\documentclass{emulateapj}

\newcommand{\be}{\begin{equation}}
\newcommand{\ee}{\end{equation}}

\def\farcs{\hbox{$.\!\!^{\prime\prime}$}}
\def\farcm{\hbox{$.\!\!^{\prime}$}}

\newcommand{\gapr}{\raisebox{-.6ex}{\mbox{
$\stackrel{>}{\mbox{\scriptsize$\sim$}}\:$}}}

\begin{document}

\title{
X-ray emission from the planet pulsar B1257+12}

\author{
G.\ G.\ Pavlov, O.\ Kargaltsev, G.\ P.\ Garmire, A.\ Wolszczan
}
\affil{The Pennsylvania State University, 525 Davey Lab,
University Park, PA 16802, USA} \email{pavlov@astro.psu.edu}

\begin{abstract}
We report the detection of the millisecond pulsar B1257+12 with the
{\sl Chandra X-ray Observatory}. In a 20 ks exposure we detected 25
photons from the pulsar, with energies between 0.4 and 2.0 keV, 
corresponding to the flux $F_{\rm X}=(4.4\pm 0.9)\times
10^{-15}$ ergs s$^{-1}$ cm$^{-2}$ in this energy range. The X-ray
spectrum can be described by a power-law model with photon index
$\Gamma \approx 2.8$
and luminosity $L_{\rm X} \approx 2.5\times 10^{29}$  ergs
s$^{-1}$ in the 0.3--8 keV band, for a plausible distance of 500 pc
and hydrogen column density $N_{\rm H}=3\times 10^{20}$ cm$^{-2}$.
Alternatively, the spectrum can be fitted by a blackbody model with
$kT\approx 0.22$ keV and projected emitting area $\sim 2000$ m$^2$.
If the thermal X-rays are emitted from two symmetric polar caps, the
bolometric luminosity of the two caps is
$2 L_{\rm bol}\sim 3\times 10^{29}$ ergs s$^{-1}$.
We compared our results with the data on other 30 millisecond pulsars
observed in X-rays and found that
the apparent X-ray efficiency of
PSR B1257+12, $L_X/\dot{E} \sim 3\times 10^{-5}$ for $d=500$ pc, is lower than
those of most of millisecond pulsars. This might be explained by 
an unfavorable orientation of the X-ray pulsar beam if the
radiation is magnetospheric, or by strong asymmetry of polar caps
if the radiation is thermal
(e.g., one of the polar caps 
is much brighter than the other and
remains invisible for most part of the pulsar period).
Alternatively, it could be attributed to
absorption of X-rays in circumpulsar matter, such as a flaring debris
disk left over after formation of the planetary system around the pulsar.
\end{abstract}
%%%%%%%%%%%%%%%%%%%%%%%%%%%
%%%%%%%%%%%%%%%%%%%%%%%%%%%%%
\section{Introduction}
%%%%%%%%%%%%%%%%%%%%%%%%%%%%%%
Most of millisecond radio pulsars (MSPs) reside 
in binary systems, usually with a white
dwarf companion. It is generally believed
that they have been spun up to millisecond
periods by accretion from their companions. 
About 20\%
of the known MSPs are solitary pulsars
(Lorimer 2005\footnote{See {\tt http://www.livingreviews.org/lrr-2005-7}}), 
which either lost their binary
companions or were born with short periods and low magnetic fields.
Both binary and solitary millisecond pulsars can emit X-rays from their
magnetospheres. Similar to ``ordinary'' (non-recycled) pulsars,
this magnetospheric X-ray emission is usually characterized
by a power-law (PL) spectrum with photon indices $\Gamma \approx 1$--2
and sharp X-ray pulsations (e.g., Becker \& Pavlov 2001; Zavlin 2006ab). 
In addition
to the magnetospheric emission, MSPs emit thermal X-rays from hot polar
caps.
The polar cap emission is characterized by a soft X-ray
spectrum, which resembles a blackbody (BB)
spectrum with temperatures $\sim 1$--3 MK, and smoother
pulsations. Both thermal and magnetospheric luminosities are small fractions,
$\sim 10^{-4}$--$10^{-2.5}$, of the spin-down power $\dot{E}$;
the thermal emission
apparently dominates
in solitary MSPs with
$\dot{E} 
\lesssim 10^{35}$ ergs s$^{-1}$.

Of particular interest among the solitary millisecond pulsars
is
PSR B1257+12
(also known as PSR J1300+1240; we will call it B1257 hereafter),
which hosts a planetary system
comprised of the first extrasolar planets discovered
(Wolszczan \& Frail 1992).
The pulsar's period, period derivative, and dispersion measure are $P=6.22$ ms,
$\dot{P}_{\rm obs} = 1.14\times 10^{-19}$ s s$^{-1}$, 
and DM = 10.2 cm$^{-3}$ pc,
respectively.
The %dispersion-measure
DM-based distance estimates are 620 and 450 pc
for the Galactic electron density distibution models by Taylor \& Cordes (1993) and
Cordes \& Lazio (2002), respectively, with nominal uncertainties of 20\%.
Wolszczan et al.\ (2000) reported a distance of $\sim$800 pc
from a timing parallax measurement, but this value is subject to
large errors because of strong correlations with the other timing parameters.
We will scale the distance to 500 pc below.

B1257 shows a substantial proper motion,
$46.4\pm 0.1$ and $-82.2\pm 0.2$ mas yr$^{-1}$
 in the right ascension and declination,
respectively (Wolszczan et al.\ 2000),
corresponding to the transverse velocity,
$v_\perp =
224\, d_{500}$ km s$^{-1}$, unusually large for a millisecond
pulsar.  Therefore, the period derivative should be
corrected for the
effect of pulsar's transverse
motion (Shklovskii 1970): $\dot{P} = \dot{P}_{\rm obs} - (v_\perp^2P/dc) =
(11.42 -
6.74\, d_{500})\times 10^{-20}$.
This distance-dependent correction increases the
inferred
spin-down age, $\tau = 0.863\, (1-
0.590 d_{500})^{-1}$ Gyr,
and reduces the inferred spin-down power,
% and magnetic field,
$\dot{E} = 1.88\times 10^{34}\, (1-0.590 d_{500})$ ergs s$^{-1}$,
and magnetic field,
$B=8.53\times 10^8\, (1-0.590 d_{500})^{1/2}$ G.
Notice that allowance for this effect puts an upper limit
$d < 848$ pc, on the
the distance
to B1257.

One can expect that B1257, just as other MSPs, emits
X-rays generated
in the pulsar's magnetosphere and/or hot polar caps, and
studying its X-ray
emission can be used, together with the data on other MSPs,
to understand how the polar cap and magnetosphere  properties
depend on pulsar parameters. On the other hand,
it is possible that some kind of debris in the pulsar's planetary system
can lead to additional absorption of X-rays close to the pulsar, which
might allow one to establish the presence of such debris and study their
properties. With this in mind, we proposed to observe B1257 with the
{\sl Chandra} X-ray Observatory. We describe the observation and its results
in \S2 and \S3, and discuss some implications in \S4.
%%%%%%%%%%%%%%%%%%%%%%%%%%%%%%%%%%%%%%%%%%%%
\section{Observations}
%%%%%%%%%%%%%%%%%%%%%%%%%%%%%%%%%%%%%%%%

We observed B1257 with the Advanced CCD Imaging Spectrometer (ACIS)
aboard {\sl Chandra} on 2005 May 22 (start time
53,512.220183 MJD)
for 20.05 ks (19,797 s effective exposure time).
The observation was carried out in Very Faint mode, and the pulsar was
imaged on ACIS-S3 chip with a standard Y offset of $-0\farcs33$.
The detector was operated in Full Frame mode, which provides time
resolution of 3.2 seconds. The data were reduced using the Chandra
Interactive Analysis of Observations (CIAO) software (ver.\ 3.3;
CALDB 3.2.1).
 For the analysis, we used the standard grade filtering
and restricted the energy range to 0.3--8 keV.

\section{X-ray image and spectrum}

Figure 1 shows the ACIS-S3 image of the field around B1257, with an
X-ray source
centered at ${\rm R.A.}=13^{\rm h}00^{\rm
m}03\fs10$, ${\rm decl.}=+12^{\circ}40' 56\farcs0$
(J2000).
The uncertainty of this position, 0\farcs6 at the 90\% confidence
level, is mainly caused by errors
in absolute {\sl Chandra} astrometry.
Since
it differs by only
0\farcs3 from the radio position, ${\rm
R.A.}=13^{\rm h}00^{\rm m}03\fs0810$, ${\rm decl.} =+12^{\circ}40'
55\farcs875$ for the epoch of the {\sl Chandra} observation,
we conclude with confidence that we detected the X-ray emission from
B1257. The distribution of source counts in the ACIS image is
consistent with that of a point source.

%\clearpage
\begin{figure}
 \centering
\includegraphics[width=
3.4in,angle=0]{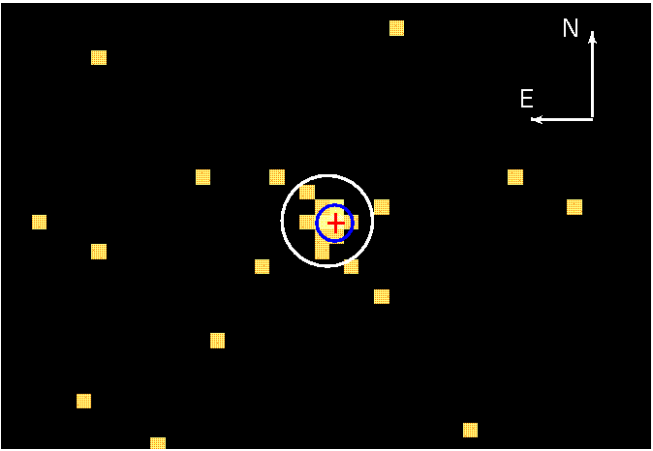}
\caption{ ACIS-S3 image of B1257+12.
The cross 
marks the radio position of PSR B1257+12, whose uncertainty is negligible
on this scale. The 0\farcs6 radius of the smaller (blue) circle corresponds
to the 90\% uncertainty of the {\sl Chandra} pointing. The larger (white)
circle of 1\farcs5 radius encompasses the area from which the photons
used for the spectral analysis were extracted.
 }
\end{figure}
%\clearpage

To measure the flux and the spectrum of the pulsar,
we chose a circular aperture of 1\farcs5 radius (about 3 ACIS pixels),
which contains 95\% encircled energy fraction.
Using
the CIAO {\tt psextract} task,
we found 25 events within this aperture.
Scaling the background (495 counts in an annulus of 14,877 pixel
area) to the source aperture, we found the average background
contribution of 0.94 counts. The background-subtracted,
aperture-corrected source flux is 
$F_{\rm X}= (4.4\pm 0.9)\times 10^{-15}$ 
ergs s$^{-1}$ cm$^{-2}$ (the errors here and below
are at the 68\% confidence level for one interesting parameter),
in the 0.4--2.0 keV band that
includes the energies of all the 25 events detected (see Fig.\ 2).

In Figure 2
 we
show the distribution of arrival times for the 25 detected photons
over the duration of the observation.
The distribution of arrival times does not show any statistically
significant deviations from the Poisson statistics.

We fitted the spectrum in the 0.3--8 keV range with the absorbed BB and PL 
models. 
Since grouping of the detected 25 counts in energy bins
would result in small numbers of counts per
bin, using the standard $\chi^2$ minimization technique would introduce
a significant bias in the deduced model parameters (Cash 1979; Nousek \& Shue
1989). Therefore, we have to use
the C-statistic
(implemented in XSPEC, ver.\ 11.3.2), without energy binning.
To obtain constrained
fits, we had to freeze the hydrogen column density, $N_{\rm H}$. The
pulsar's dispersion measure,
${\rm DM} =10.2$ cm$^{-3}$ pc,
corresponds to $N_{\rm H}  
\simeq 3\times 10^{20}$ cm$^{-2}$
(assuming a 10\% ISM ionization), which we adopt in our fits
(see Table 1 and Fig.\ 3).
%\clearpage
%\hoffset=-0.5in
\begin{figure}
% \centering
\hspace{-0.3in}
\includegraphics[width=2.8in,angle=90]{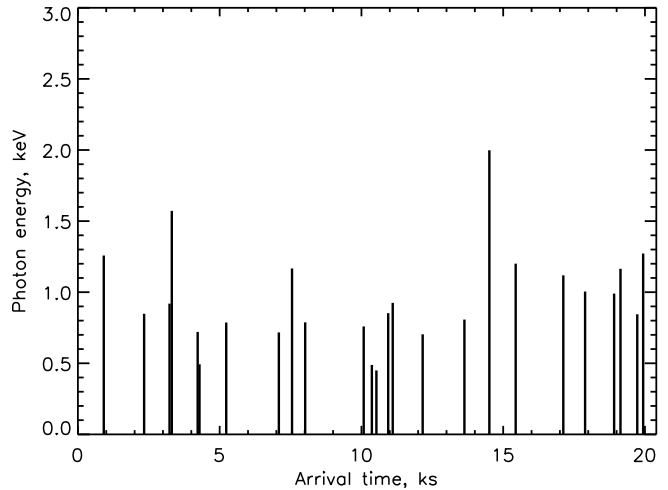}
\caption{
Energies and arrival times for 25 photons from the 1\farcs5 radius
aperture around B1257.
 }
\end{figure}

%\clearpage
The PL fit gives a photon index
$\Gamma=2.75 \pm 0.35$ and normalization constant ${\cal N} =
(1.7\pm 0.3)\times 10^{-6}$ photons cm$^{-2}$ s$^{-1}$ at 1 keV,
corresponding to
the luminosity $L_X=4\pi d^2 F_X^{\rm unabs} \simeq(2.5\pm0.5)\times
10^{29}d_{500}^2$ ergs s$^{-1}$ in the 0.3--8 keV band, for
isotropic emission.
The apparent temperature, 
$T=2.5\pm 0.3$ MK,
 and the apparent projected area of the emitting region,
$A_{\perp a}= 2.1^{+1.9}_{-0.9}\times 10^3\, d_{500}^2$ m$^2$,
obtained from the BB fit are strongly correlated (see Fig.\ 3),
which explains the large uncertainties of these parameters.
Since $A_{\perp a}$ is 5 orders of magnitude smaller than the 
assumed
projected area of neutron star (NS) surface,
$\pi R_{\rm NS}^2$,
with $R_{\rm NS}\approx 10^4$ m$^2$,
 such thermal radiation could originate only from small heated spots.
If there are two identical spots (polar caps) at star's magnetic poles,
the radius and the bolometric luminosity of each of the spots are
$R = (A_{\perp a}/f\pi)^{1/2} \sim 26\, d_{500} f^{-1/2}$ m
and $L_{\rm bol} = \sigma (T_a/g_r)^4 (A_{\perp a}/f) \approx
4.6\times 10^{28} d_{500}^{2} f^{-1} g_r^{-4}$ ergs s$^{-1}$, where
the geometrical correction factor $f\leq 1$
depends on the angles $\zeta$ (between the
line of sight and the spin axis) and $\alpha$ (between the magnetic
and spin axes) as well as the gravitational redshift factor $g_r =
(1-R_g/R_{\rm NS})^{1/2}$ ($R_g = 2953\, M_{\rm NS}/M_\odot$ m is
the gravitational radius).
Because of a substantial scatter and a shallow swing of the polarization 
position angle
within the pulse profile in radio polarimetric observations
(Xilouris et al.\ 1998), the values of $\zeta$ and $\alpha$ remain highly
uncertain for B1257. However, thanks to the GR effect of
bending photon trajectories
in the NS gravitational field, the factor $f$ varies in a relatively
narrow range:
 $2R_g/R \lesssim f \leq 1$ (or 
$0.83 \lesssim f \leq 1$ for 
$M_{\rm NS}=1.4 M_\odot$
and $R_{\rm NS}=10^4$ m, i.e. $g_r=0.766$), 
where the lower limit (corresponding 
to $\alpha=0$, $\zeta=90^\circ$ or $\zeta=0$, $\alpha=90^\circ$) is
estimated in the approximation outlined in Appendix of
Zavlin, Shibanov, \& Pavlov (1995).
The BB model fits the spectrum better than
the PL model (the C-statistic values are 77 and 84, respectively, for 524 degrees of
freedom), but the PL fit cannot be rejected based on statistical arguments.

%\clearpage
\begin{figure}[]
% \centering
\hspace{-0.3in}
 \vbox{
\includegraphics[width=
2.8in,angle=90]{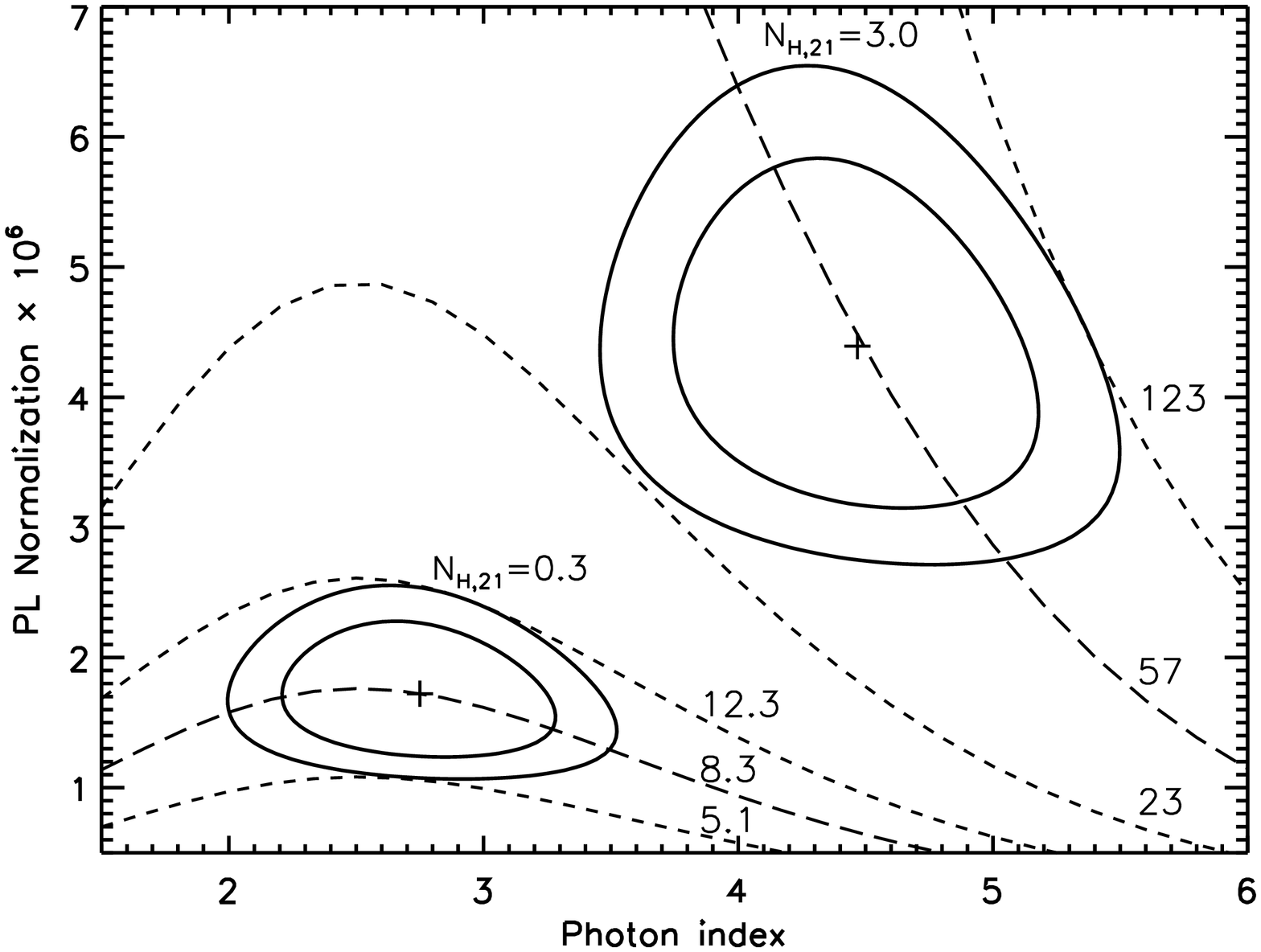}
\vspace{-0.0cm}
\includegraphics[width=
2.8in,angle=90]{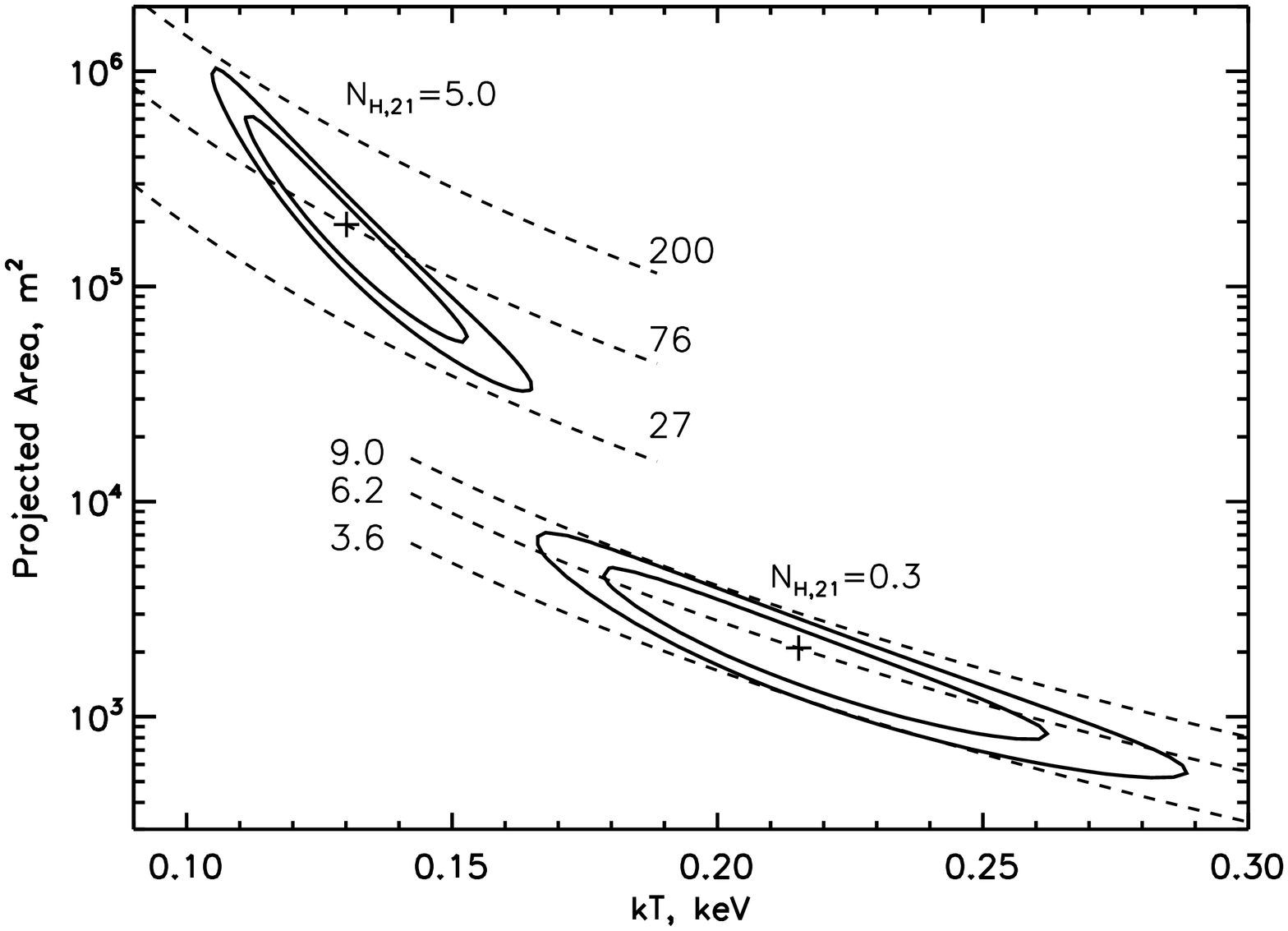}}
\caption{ Confidence contours (68\% and 90\%) for the PL
({\em top}) and BB ({\em bottom}) fits to the spectrum of B1257,
for fixed $N_{\rm H}$ values ($N_{\rm H} =
0.3\times
10^{21}$ cm$^{-2}$ corresponds to the pulsar's dispersion measure,
while $N_{\rm H} = 3\times 10^{21}$ and $5\times 10^{21}$ cm$^{-2}$ were chosen
to obtain X-ray efficiency of B1257 similar to those of other pulsars;
see \S4.2).
The PL
normalization is in units of $10^{-6}$ photons cm$^{-2}$ s$^{-1}$
keV$^{-1}$ at 1 keV.
The BB normalization
is the projected emitting area
in units of m$^2$, for $d = 500$ pc.
The dashed lines are the lines of constant unabsorbed flux in the 0.3--8 keV 
band in the top
panel,
and constant bolometric flux in the bottom panel,
both in units of 10$^{-15}$ ergs cm$^{-2}$ s$^{-1}$.
}
\end{figure}
%\clearpage

\section{Discussion}

X-ray emission from a solitary MSP
(or from an MSP in a wide binary)
can be produced
in the pulsar's magnetosphere and/or hot polar caps,
while an additional nonthermal component, emission from an intrabinary
shock, can become important if the MSP is in a close binary system.
The thermal (polar cap) and nonthermal (magnetospheric and shock) contributions
 can be distinguished by the shape of the X-ray
spectrum.
If the observed emission is
predominantly nonthermal, the spectrum is best described by a PL, which is
particularly hard when magnetospheric emission dominates
(e.g., PSR J0218+4232, B1821$-$24, B1937+20, for which $\Gamma =1$--2; 
see Table 2).
On the contrary, the spectrum
of the thermal polar
cap emission is quite soft; when fit with a PL model, it shows
substantially larger photon indices, $\Gamma =3$--5.
Thermal emission is not seen in younger
($\tau \lesssim 0.5$ Gyr), powerful ($\dot{E} \gtrsim 10^{35}$ ergs s$^{-1}$) MSPs,
while it dominates in
solitary MSPs with $\dot{E} \lesssim 10^{35}$ ergs s$^{-1}$.
Observations of
nearby thermally-emitting MSPs,
such as PSR J0437--4715 (Zavlin et al.\ 2002;
Zavlin 2006a,b), actually show both thermal and nonthermal components,
the former dominating at lower energies, $\lesssim 2$ keV.

Since the spin-down power of B1257 is $\lesssim 10^{34}$ erg s$^{-1}$,
and its spectrum is quite soft
($\Gamma \sim 3$) and detected only below 2 keV, one can expect
that the detected emission is mostly thermal emission from polar caps.
If this is the case, the best-fit BB temperature, $kT_a\approx 0.2$ keV,
is about the same as found in other MSPs (e.g., PSR J0437--4715, J2124--3358,
and J1024--0719).
Although the projected emitting area, $A_{\perp a} \sim 2\times 
10^3$ m$^2$,
is rather uncertain, it
is much smaller than the 
conventional polar cap area, $A_{\rm pc} = 2\pi^2 R_{\rm NS}^3/(Pc) =
1.05\times 10^7$ m$^2$ (for $R_{\rm NS}=
10^4$ m) predicted by simple pulsar models,
 and, correspondingly,
the best-fit BB polar cap radius, $R
\sim 30$ m,
is much smaller than the
predicted radius, $R_{\rm pc} = 
1800\, {\rm m}\sim 60 R$.
Similar (albeit not so strong) discrepancies 
have been
found for BB fits of other pulsar spectra, including ordinary old pulsars
possibly showing thermal polar cap emission
(e.g., Zavlin \& Pavlov 2004; Kargaltsev, Pavlov, \& Garmire 2006a).
For many pulsars, the discrepancy
 can be alleviated assuming that the polar cap is covered by a hydrogen
or helium atmosphere, in which case the effective temperature would be a factor of 2 lower,
and the radius a factor of 3--10 larger
(Zavlin, Pavlov, \& Shibanov 1996). However, the discrepancy is too high for B1257
to be explained in this way. Another explanation for such a discrepancy was suggested
by Zhang, Sanwal, \& Pavlov (2005), who analyzed the X-ray emission from
the ordinary drifting pulsar B0943+16
and suggested that only a small fraction of the polar cap
surface, associated with footprints of sparks produced by intermittent breakdowns
of an ``inner gap'' above the polar cap, is hot enough to emit X-rays.
However, the ``filling factor'',
$A/A_{\rm pc} \sim 2\times 10^{-4}$, is a factor of 100 smaller
for B1257 than for B0943+16, and it is currently unclear whether or not this hypothesis
is applicable to MSPs.
The area would become much larger
if the actual $N_{\rm H}$ is substantially larger than the $3\times 10^{20}\, {\rm cm}^{-2}$
estimated from the pulsar's dispersion measure 
(e.g., $A_{\perp, a}\sim 
2\times 10^5$ m$^2$ for
$N_{\rm H} = 3\times 10^{21}\,{\rm cm}^{-2}$, as demonstrated in Fig.\ 3), but
such large $N_{\rm H}$ values would require an additional absorber, such as a dust cloud
near the pulsar (see \S4.2), because they strongly exceed the total Galactic $N_{\rm H}$
in the direction to B1257 (e.g., $N_{\rm HI}=2.2\times 10^{20}\,{\rm cm}^{-2}$; Dickey \&
Lockman 1990).

The magnetospheric interpretation of the X-ray emission from B1257
looks less plausible than the thermal one, but we cannot firmly rule it out
with the small number of counts detected. Although the best-fit power-law is
unusually soft for magnetospheric emission,
the uncertainty of photon index
allows smaller values of $\Gamma$ (see Fig.\ 3),
possibly compatible with the magnetospheric
interpretation.

\subsection{Intrinsically underluminous millisecond pulsar?}

Whether the detected X-ray emission is
thermal or nonthermal,
the B1257's apparent luminosity is very low in comparison with 
other MSPs detected
in X-rays, and its apparent X-ray efficiency\footnote{We emphasize
that this low efficiency is obtained assuming $N_{\rm H}=3\times 10^{20}$
cm$^{-2}$, estimated from the pulsar's dispersion measure. The inferred 
luminosity and efficiency would be higher if a larger $N_{\rm H}$ value
is assumed, but they still would be lower than typical ones as long as
the X-ray absorption is caused by the ISM.},
$\eta \equiv L_{\rm X}/\dot{E} \approx
1.3\times 10^{-5} d_{500}^2(1-0.59 d_{500})^{-1}$, is smaller than the typical values,
$\eta \sim 10^{-4}$--$10^{-2.5}$.
This is demonstrated in Figure 4, which shows the 0.1--10 keV luminosities and spin-down powers
for 31 MSPs
(we chose the 0.1--10 keV band following Bogdanov et al.\ 2006,
whose results on 14 MSPs in the globular cluster 47 Tuc are included 
in Fig.\ 4).
The X-ray luminosities and the corrections of spin-down power for the Shklovskii effect (see \S1)
were calculated using the distances and transverse velocities listed in Table 2.
To make the picture more legible, we
 chose not to show the $L_{\rm X}$ and $\dot{E}$ uncertainties, except for B1257
(a comprehensive discussion on the $L_{\rm X}$ uncertainties is presented by Bogdanov et al.\ 2006).
For the MSPs
showing predominantly nonthermal X-ray emission
(blue squares and green triangles in Fig.\ 4), we plot the ``isotropic
luminosities'', $L_{\rm X} = 4\pi d^2 F^{\rm unabs}_{\rm X}$, which can be higher or
lower than the actual luminosities, depending on the (unknown) intrinsic angular
distribution of the pulsar's radiation.
For the MSPs
whose X-ray emission is most likely thermal (red circles in Fig.\ 4),
we used the ``equivalent sphere luminosities'' (corrected for the 
general relativity effects
assuming $g_r=0.766$) instead of the ``true'' polar cap luminosities, because
the angles $\zeta$ and $\alpha$
needed for calculation of the latters (see \S3)
are not known for most of the MSPs.
For instance, the equivalent
sphere bolometric BB luminosity is
$L_{\rm bol}^{\rm es} \equiv 4 (A_{\perp a} g_r^2)\sigma (T_a/g_r)^4 = (2fg_r^2)(2L_{\rm bol})$,
where $2 L_{\rm bol}$ is the luminosity of two polar caps. The factor $2fg_r^2$ is not very different from 1
for any $\zeta$ and $\alpha$ at the expected values of $g_r$;
e.g., 
it varies between 0.97 (at $\zeta=0$, $\alpha=90^\circ$) and 1.17 (at $\zeta=\alpha=0$) at $g_r=0.766$.

%\clearpage
Since both $L_{\rm X}$ and $\dot{E}$ depend on distance, which has not been accurately measured
for B1257, the measured X-ray flux (with account for measurement uncertainties) corresponds to a strip in
the $\dot{E}$-$L_{\rm X}$ plane, shown in Figure 4 for the $d=300$--800 pc 
range (the dashed curve
corresponds to the best fit). We see that for the most plausible distances,
$d\lesssim 750$ pc, not only the B1257's X-ray luminosity is lower than for any MSP detected in X-rays,
but also its X-ray efficiency is lower,
except may be
PSR J0034--0534 (\#2 in Fig.\ 4), for which we know only an upper limit on $\dot{E}$ because its proper
motion has not been measured. Only for distances approaching the limiting value of 848 pc
(see \S1), $\eta$ becomes similar to those of the majority of MSPs (but the luminosity,
$L_{\rm X} \to 9\times 10^{29}$ ergs s$^{-1}$, still remains lower than those of most of the other MSPs).
If the observed B1257's radiation is magnetospheric,
the low apparent luminosity might be explained by an unfavorable direction of 
the pulsar beams (i.e., the actual
luminosity of B1257 being larger than the isotropic luminosity).
However, it cannot be explained under the
more plausible
assumption that it is thermal radiation from
two
symmetric, isotropically emitting
polar caps.
Under this assumption, the maximal value of $2L_{\rm bol}/L_{\rm bol}^{\rm es} = (2f g_r^2)^{-1}$
does not exceed 1.03
for $g_r=0.766$ (see above), i.e. the maximum luminosity of two polar caps
corresponding to the observed flux is only 3\% higher than the equivalent sphere luminosity
plotted in Figure 4. 
To get a larger intrinsic luminosity, one
might speculate that the caps are very asymmetric (e.g., one of them
is much brighter than the other and is invisible for most part of pulsar period,
which can occur only when both $\zeta$ and $\alpha$ are very small.
%in contradiction with the radio polarimetric data). 
Also, the polar cap emission
can be anisotropic, contrary to BB emission, 
because of the limb-darkening effect in the
NS atmosphere, but this anisotropy is not strong at the relatively low magnetic fields
of MSPs (see Zavlin et al.\ 1996). Thus, although there are ways to derive larger
luminosity and efficiency from the same observed flux, these quantities are not expected to be much
larger than our current estimates, at least if the observed emission is indeed thermal.

Among the MSPs detected in X-rays, of particular interest is
PSR B1620--26 (\#11 in Fig.\ 4), a member of a triple system that likely
contains a Jupiter-mass planet on a distant orbit around the inner pulsar + white dwarf binary
($P_{\rm bin} = 191$ d, $P_{\rm planet}\sim 100$ yr;
Sigurdsson \& Thorsett 2005, and references therein).
The globular cluster M4, which hosts
PSR B1620--26, was observed with {\sl Chandra} ACIS on 2000 June 30 for 25.8 ks
(Bassa et al.\ 2004).
Since the description of the results on B1620--26 is very sketchy in that paper,
we reanalyzed the data using the same approach as for B1257 (\S3).
The pulsar was detected 1\farcm2 off-axis, and its image looks slightly extended
in the north-south direction. In an elliptical aperture with major and minor
axes of 
2\farcs7 and 1\farcs7 we found 21 photons in the 0.3--8 keV band
and measured the flux $F_{\rm X}=(4.2\pm 1.0)\times 10^{-15}$ ergs cm$^{-2}$ s$^{-1}$.
We fit the source spectrum with the PL and BB models at fixed
$N_{\rm H}=2.36\times 10^{21}$ cm$^{-2}$.
The PL fit gives $\Gamma=2.3\pm 0.4$ 
and $F_{\rm X}^{\rm unabs} \approx 7.7\times 10^{-15}$ ergs cm$^{-2}$ s$^{-1}$,
corresponding to the isotropic luminosity of $2.8\times 10^{30}$ ergs s$^{-1}$ in
the 0.3--8 keV band ($L_{\rm X} = 4.6\times 10^{30}$ ergs s$^{-1}$ in the 0.1--10 keV
band) at the distance  of 1.73 kpc to the globular cluster.
The BB fit gives $kT_a = 0.45\pm 0.08$ keV,
$A_{\rm \perp a}= 1.0^{+0.9}_{-0.5}\times 10^{3}$ m$^2$,
 and $L_{\rm bol}^{\rm es}
\approx 3.0\times 10^{30}$ ergs s$^{-1}$. Similar to B1257, both fits are statistically
acceptable and give about the same X-ray luminosity. The apparent BB temperature
is higher than for the other thermally emitting MSPs, which indicates that
we possibly detect both thermal and nonthermal radiation, perhaps even some contribution from
a pulsar wind nebula, as indicated by the possible extension of the source image.
Regardless of the emission mechanism, the luminosty of this MSP is substantially
higher than that of B1257. Unfortunately, we 
do not know its intrinsic
spin-down power (hence, X-ray efficiency)
 because of poor knowledge of parameters of the putative
planet
that strongly contributes to the observed value of $\dot{P}$.
In Figure 4,
we plotted the point corresponding to B1620--26 at $\dot{E} = 0.01 \dot{E}_{\rm obs}$,
following the assumption by Thorsett et al.\ (1999), and showed the $\dot{E}$ uncertainty
by the double-sided horizontal arrow. Notice that if $\dot{E}$ is indeed so small, then B1620--26
is the most efficient X-ray emitter ($\eta\sim 10^{-2}$) among the 31 MSPs,
opposite to 
the other planet pulsar
B1257. We should remember, however, that the planet in the
B1620$-$26 system is quite different in properties and history from
the B1257 planetary system.

\subsection{Absorbed by matter orbiting the pulsar?}

The planets around B1257 were likely formed from a disrupted
or ablated stellar companion that had possibly provided the matter to spin up the
pulsar to its millisecond period
(Phinney \& Hansen 1993; Podsiadlowski 1993).
Some material left over the planet formation (asteroids, meteoroids, dust)
can still rotate around the pulsar (e.g., Bryden et al.\ 2006;  
Cordes \& Shannon 2006). 
Absorption of the B1257's radiation by such 
circumpusar material might
explain the relatively low observed X-ray flux.

To explore this possibility, we assume that X-rays pass through a cloud 
that contains some grains or rocks.
For optically thick grains
(radius $a \gg 1\,\mu$m at $E\sim 1$ keV),
the grain cross section, $\sigma_g \sim \pi a^2$, is independent of photon energy,
so the absorption by grains does not affect the shape of X-ray spectrum.
To obtain the intrinsic B1257's luminosity and efficiency similar to those of
other MSPs,
the optical depth $\tau$ of the intervening cloud should be in a range of 2--5,
along the line of sight, corresponding to the column density
$N_g \sim \tau (\pi a^2)^{-1}$ and number density
$n_g\sim 2\times 10^{-14} \tau a^{-2} l^{-1}$ cm$^{-3}$,
where $a$ is the grain radius in units of cm, and
$l$ is is the path length through the cloud in AU.
The mass of the putative cloud
 depends on its geometry and location with respect
to the line of sight, as well as on size and composition of grains, all of which
are unknown. It can be scaled as
\be
M_{\rm cl} \sim 3\times 10^{26}\tau (a/l) \rho V_{\rm cl}\,\,\,{\rm g}\,,
\ee
where $\rho$ is the mass density of the grain material in g\,cm$^{-3}$, and $V_{\rm cl}$ the cloud's volume
in units of AU$^3$.
Thus, at $\tau\sim 3$, $\rho\sim 2$ g cm$^{-3}$, and a characteristic cloud size of 1 AU,
we obtain $M_{\rm cl} \sim (0.003$--$300) M_\earth$ for $a=0.01$--1000 cm.
If the absorption of X-rays is caused by large bodies,
the mass of the cloud becomes uncomfortably large
(e.g., in comparison with the total mass of B1257's planets, $\approx 8 M_\earth$),
especially if the mass distribution is concentrated towards the planets orbital plane.
Since the orbital plane is inclined by $40^\circ$ to the line of sight (Konacki \& Wolszczan 2003),
only a periphery of such a distribution (a flaring disk)
 can contribute to the absorption,
and this would imply $V_{\rm cl} \gg 1$.
On the other hand,
distant orbits of circumpulsar matter can be strongly inclined with respect to the planets orbital
plane, so that one might imagine an azimuthally nonuniform  belt of matter on an orbit
partially eclipsing the pulsar.
An argument against this hypothesis is that
grains/rocks are expected to have been evaporated/ablated by the pulsar wind
unless they are very large (e.g., $\gtrsim 100$ m in size; see Miller \& Hamilton 2001)
or replenished by collisions of larger bodies
with a rate exceeding the rate of evaporation/ablation.

Absorption of X-rays by very small, optically thin grains
is virtually indistinguishable
from absorption by the ISM gas
(Wilms et al.\ 2000, and references therein). Therefore, its effect on the absorbed
spectrum can be crudely modeled by increasing $N_{\rm H}$ in the ISM absorption
models.
To explore this possibility, we fitted the X-ray spectrum of B1257 with
the same PL and BB models as previously but assuming larger (fixed) values for
$N_{\rm H}$. We show the parameter confidence contours in Figure 3 for $N_{\rm H} = 3$ and
$5 \times 10^{21}$ cm$^{-2}$ for the PL and BB fits respectively 
(these values provide a factor of $\sim$10
higher luminosities for the two models). The PL fit yields a very large photon
index, $\Gamma \sim 4$--5, suggesting that the spectrum is, in fact, thermal.
The BB fit gives a slightly lower temperature, $T_a = 1.51\pm 0.16$ MK and
a much larger emitting area, $A_{\perp a} = 
1.9^{+2.6}_{-0.9}\times 10^5$ m$^2$
($\sim 0.02 A_{\rm pc}$),
corresponding to a luminosity
$L_{\rm bol}\sim 5.6\times 10^{29}d_{500}^2 f^{-1}g_r^{-4}$ ergs s$^{-1}$.
A factor of 10 higher X-ray efficiency obtained from this fit is similar to
those of most of MSPs.
The larger $N_{\rm H}$ required for this increase can be used to estimate
the mass of the intervening cloud, 
\begin{equation}
M_{\rm cl} \sim 2\times 10^{24} \xi V_{\rm cl} l^{-1}\,{\rm g}
\end{equation}
where $\xi
\sim Z_{\rm st}/Z$ is a factor depending on element abundances in the grains
($Z_{\rm st}\approx 0.02$ 
is the `standard' metallicity used in the absorption model, and $Z$ is the actual
metallicity in the grains).
We see that a much lower mass, $M_{\rm cl} \sim 3\times 10^{-4} \xi M_\earth$ for
a 1 AU characteristic cloud size, is needed to explain the low X-ray efficiency
of B1257 by absorption in a circumpulsar dust of microparticles.
However, 
such small particles could be blown out
from the pulsar's vicinity by the pulsar wind and radiative pressure
(especially at earlier epochs when the spindown liminosity was higher),
and they could be swept out from the outskirts of
 the planetary system by the ram pressure
of the oncomimg medium. Therefore, to explain the low apparent X-ray 
luminosity and efficiency as due to absorption by small particles,
we have to assume that the particles are being replenished
by collisions of larger bodies in the putative asteroid belt.

Some additional information on the circumpulsar matter can be provided by 
infrared (IR) observations.
Since such matter is being heated by the pulsar's radiation (photons and pulsar wind),
it should emit IR radiation whose spectrum depends on grain's temperature and composition.
So far, searches for such emission at $\lambda \sim 10$--1000 $\mu$m have yielded only upper limits
(Bryden et al.\ 2006, and references therein), which, however, can be used to put some constraints
on the circumpulsar matter properties. For instance, from the upper limit on spectral flux,
$F_\nu < 45\,\mu$Jy at $\lambda=24\,\mu$m (Bryden et al.\ 2006),
we obtain an upper limit,
$A<1.5\times 10^{23} [\exp(600\,{\rm K}/T)-1] q_{24}^{-1} d_{500}^2$ cm$^2$,
on the emitting area of grains in the cloud ($q_{24}$ is the emission efficiency
$q_\lambda$ at $\lambda=24\,\mu$m).
If the cloud is optically thin in IR, this limit
translates into $M_{\rm cl}\sim A a \rho/3 
< 0.5\times 10^{23} a \rho\, [\exp(600\,{\rm K}/T)-1] q_{24}^{-1} d_{500}^2$ g
(e.g., $M_{\rm cl} < 4\times 10^{25} a$ g for $T=100$ K, $\rho = 2$ g cm$^{-3}$,
$a \gapr 0.5\times 10^{-2}$ cm),
and, together with the estimates of cloud mass needed to explain the additional
absorption (equations [1] or [2]),
it constrains the cloud size. For instance, in the case of large grains,
we obtain
$V_{\rm cl}/l < 1.7\times 10^{-4} [\exp(600/T)-1] \tau^{-1}$ AU$^2$
(e.g., $V_{\rm cl}/l < 2.2\times 10^{-2}$ AU$^2$ for $T=100$ K, $\tau=3$).
Since such estimates strongly depend on
the unknown temperature, it would be very useful to obtain better constraints
on the temperature and emitting area from deeper IR observations.

Overall, the hypothesis that the low apparent X-ray efficiency is caused by
absorption of X-rays in circumpulsar matter does not look unreasonable
at this point.
However, it implies a rather narrow range of optical depths
(and, in the case of large
grains, rather large masses of absorbing matter)
to make the
B1257's X-ray efficiency consistent with those of other MSPs.
In addition to deep IR observations,
a possible way to confirm or reject this interpretation would be monitoring
of the pulsar's X-ray emission.
If the X-ray flux shows substantial variations (e.g., on a year timescale),
it could be caused by variable absorption in an orbiting, nonuniformly
distributed circumpulsar matter.

\acknowledgements
We thank
Kiriaki Xilouris for discussions
on radio polarimetric observations of B1257,
and the anonymous referee for very useful remarks.
This work was partially supported by NASA grants NAG5-10865
and NAS8-01128 and {\sl Chandra} award SV4-74018.

\clearpage
\begin{figure}
\hspace{-0.6in}
\includegraphics[width=5.8in,angle=90]{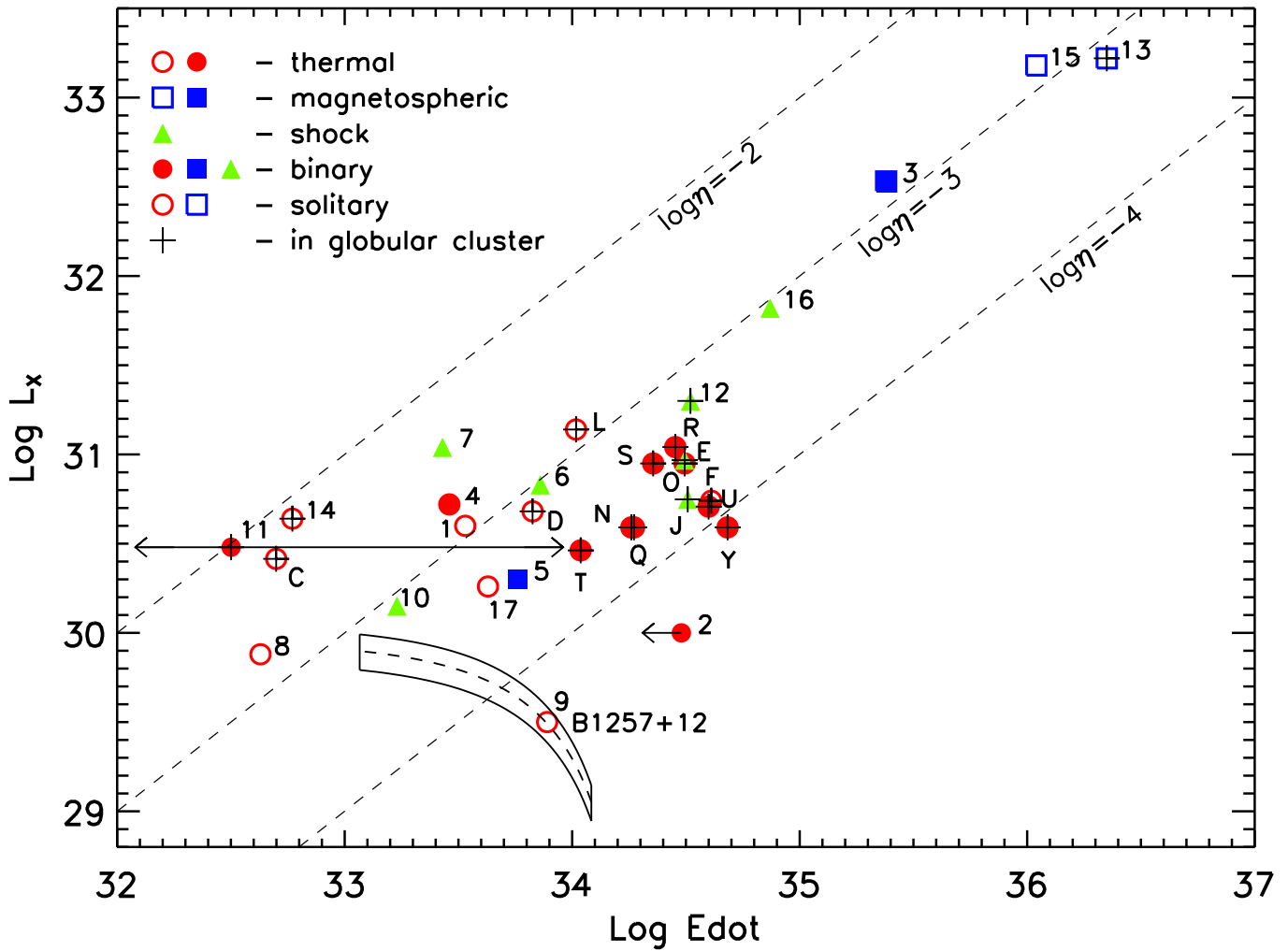}
\caption{X-ray luminosty in the 0.1--10 keV band versus intrinsic spin-down powe
r for
31 MSPs observed with {\sl Chandra} and {\sl XMM-Newton}. The numerical labels c
orrespond to
the numbers listed in first column of Table 2, while the letters
mark the MSPs in 47 Tuc (Table 4 in Bogdanov et al.\ 2006). The red circles,
blue squares, and green triangles correspond to MSPs whose X-ray emission is dom
inated by
thermal polar cap, magnetospheric, and intrabinary shock emission,
respectively. The open and filled symbols denote solitary and binary
pulsars, respectively. MSPs in globular clusters are marked by black + signs.
The double-sided horizontal arrow attached to \#11 (PSR B1620--26 in M4)
demonstrates the uncertainty of its intrinsic $\dot{E}$ (see text).
For \#2 (PSR J0034--0534), only an upper limit on $\dot{E}$ is known because
the lack of proper motion measurements makes the correction for the Shklovskii
effect impossible. The curved strip shows possible positions for B1257
in the $\dot{E}$-$L_{\rm X}$ plane for a range of distances,
300--800 pc
 (the dashed curve within the strip corresponds to the
best-fit luminosity for a given distance, and the point \#9 corresponds to $d=50
0$ pc).
Three dashed straight lines correspond to constant values of the X-ray
efficiency, $\eta = L_{\rm X}/\dot{E}$.
 }
\end{figure}

\clearpage

\begin{table}[]
\caption[]{
Fitting parameters for the PL and BB models
}
\vspace{-0.4cm}
\begin{center}
\begin{tabular}{c@{\extracolsep{0.3in}}cccccc@{\extracolsep{-0.4in}}}
\tableline\tableline Model & $N_{\rm H,20}$   &
$\mathcal{N}$\tablenotemark{a} or $A_{\perp a}$\tablenotemark{b} &
$\Gamma$ or $kT$\tablenotemark{c} & $C$/dof  & $L_{\rm X}$ or $L_{\rm bol}$\tablenotemark{d} \\
%\tableline
\colrule
      PL              &       3          &      $1.72^{+0.34}_{-0.34}$   &     $2.75^{+0.34}_{-0.36}$ & 84/524  & $2.47_{-0.48}^{+0.50}$  \\
      BB              &       3          &      $2.1^{+1.9}_{-0.9}$       &  $0.215_{-0.023}^{+0.025}$ &  $77/524$   & 
$1.84^{+0.32}_{-0.40}$  \\
      PL              &       30           &      $4.39^{+0.84}_{-0.84}$   &     $4.47^{+0.46}_{-0.46}$ & 77/524  & $16.6_{-5.1}^{+7.6}$  \\
      BB              &       50           &      $190^{+260}_{-100}$       &  $0.130_{-0.014}^{+0.015}$ &  $77/524$   & 
$22.9^{+9.2}_{-10.9}$ \\
      \tableline
\end{tabular}
\end{center}
\tablecomments{The fits are for fixed $N_{\rm H,20} \equiv N_{\rm H}/10^{20}$
cm$^{-2}$ (second column).}
\tablenotetext{a}{Spectral flux in units of $10^{-6}$ photons
cm$^{-2}$ s$^{-1}$ keV$^{-1}$ at 1 keV.} \tablenotetext{b}{Projected
area of emitting region for the BB model (in $10^{3}$ m$^2$) for
$d=500$ pc} \tablenotetext{c}{BB temperature in keV.}
\tablenotetext{d}{Unabsorbed luminosity in the 0.3--8 keV band or
apparent bolometric luminosity ($4\sigma T_a^4 A_{\perp a}$), in
units of $10^{29}$ ergs s$^{-1}$ for $d=500$ pc.}
\end{table}

%\begin{deluxetable}{cllrcrrccrrr}
\begin{table}[]
%\tabletypesize{\scriptsize} 
%\tablecaption{
\caption{Millisecond pulsars observed with {\sl Chandra} and {\sl XMM-Newton}}
\vspace{-0.4cm}
%
%\tablehead{
%\colhead{\#} &
%\colhead{Name\tablenotemark{a}} & \colhead{Type\tablenotemark{b}} & \colhead{$P$} &
%\colhead{$P_{\rm orb}$\tablenotemark{c}} & \colhead{$v_\perp$\tablenotemark{d}} &
%\colhead{$\log\dot{E}$\tablenotemark{e}} & \colhead{$d$\tablenotemark{f}} &
%\colhead{$\Gamma$\tablenotemark{g}} &
%\colhead{$\log L_{\rm X}$\tablenotemark{h}} & \colhead{$-\log \eta$\tablenotemark{i}} & \colhead{Refs.}
%}\\
\begin{center}
\begin{tabular}{c@{\extracolsep{0.1in}}llrcrrccrrr}
\colrule\colrule
\# & Name\tablenotemark{a} & Type\tablenotemark{b} & $P$ & $P_{\rm orb}$\tablenotemark{c} & $v_\perp$\tablenotemark{d} & $\log\dot{E}$\tablenotemark{e} & $d$\tablenotemark{f} & $\Gamma$\tablenotemark{g} & $\log L_{\rm X}$\tablenotemark{h} & $-\log \eta$\tablenotemark{i} & Refs. \\
%\colrule
%\startdata
    &    &     & ms   & d    & km/s &       &   kpc  &     &    &  &   \\
%\tableline 
\colrule
1 & J0030+0451 & th & 4.87 & ...  & $<$16& 33.53 &  0.30\tablenotemark{j}  & 4.7 & 30.60 & 2.93  & 1,2   \\
2 & J0034$-$0534\tablenotemark{k} & th? & 1.88 & 1.59 & ...  & $<$34.48 &  0.53 & 2.5 & 30.00  & $<$4.48 &  3,4   \\
3 & J0218+4232 & mag & 2.32 & 2.03 & 50   & 35.38 &  2.6    & 1.1 &32.53 & 2.85 & 5,6  \\
4 & J0437$-$4715 & th  & 5.76 & 5.74 & 
106 & 33.46 & 0.16\tablenotemark{j} & 4.1 & 30.72 & 2.74 & 7,8 \\
5 & J0737$-$3039A\tablenotemark{l} & mag? & 22.70& 0.10 & 10   & 33.76 &  0.48  & 3.2 & 30.30 & 3.64 & 9-12 \\
6 &J0751+1807 & sh & 3.48 & 0.26 & 32   & 33.86 & 1.12  & 1.6 & 30.83 & 3.03 &  13,14  \\
7 & J1012+5307 &sh? & 5.26 & 0.60 &
             102 & 33.43 &  0.84\tablenotemark{m} & 1.8 & 31.04 & 2.39 & 13--16    \\
8 & J1024$-$0719 & th & 5.16 & ...  & 109  & 32.63&  0.39& 3.7 & 29.88 & 2.75 & 7,8  \\
9 & {\bf B1257+12} & th & 6.22 & ...  & 224  & 33.89 &  $\sim$0.50  & 2.7 & 29.50 & 4.44 & 17,18  \\
10 & B1534+12\tablenotemark{l}   & sh? & 37.90& 0.42 & 120  & 33.23 &  1.0\tablenotemark{j}  & 3.2 & 30.15
& 3.08 & 10,19  \\
11 & B1620$-$26\tablenotemark{n} (M4) & th? & 11.07 & 191 & ... & $<$34.20 & 1.73\tablenotemark{p} & 2.4 & 30.48 &$<$3.72
 & 20,18 \\
12 & J1740$-$5340 (NGC\,6397) & sh & 3.60 & 1.34 & ...  & 34.52 & 2.55\tablenotemark{p} & 1.5 & 31.30 & 3.22 & 21,22 \\
13 & B1821$-$24 (M28) & mag  & 3.05 & ...  & 120  & 36.35 &  5.5\tablenotemark{p} & 1.2 & 33.22 & 3.13
 & 23 \\
14 & J1911$-$6000C (NGC\,6752)& th? & 5.28 & ... & ... & 32.77 &  4.1\tablenotemark{p} & 2.5 & 30.64 & 2.13 & 24 \\
15 & B1937+21  & mag & 1.56 & ...  & 14   & 36.04 &  3.55  & 1.9 & 33.18 & 2.86 & 25   \\
16 & B1957+20  & sh  & 1.61 & 0.38 & 359  & 34.87 &  2.49  & 1.9 & 31.82 & 3.05 & 26,27  \\
17 & J2124$-$3358 & th & 4.93 & ...  & 62   & 33.63 &  0.27  & 3.3 & 30.26 & 3.37 & 4,28   \\
%\enddata
\colrule
\end{tabular}
\end{center}
\tablecomments{
References: 1- Becker \& Aschenbach 2002, 2- Lommen et al.\ 2006,
3- Bailes et al.\ 1994, 4- Zavlin 2006a, 5- Navarro et al.\ 1995, 6-Webb et al.\ 2004a,
7- Hotan et al.\ 2006, 8- Zavlin et al.\ 2002, 9- Lyne, et al.\ 2004,
10- Kargaltsev et al.\ 2006b, 
11- Kramer et al.\ 2006,
12- Chatterjee et al.\ 2007,
13-Webb et al.\ 2004b, 14- Nice et al.\ 2005, 15- Lange et al.\ 2001, 16- Callanan et al.\ 1998,
17- Wolszczan et al.\ 2000, 18- this work, 19- Stairs et al.\ 2002,  20- Bassa et al.\ 2004,
21- Bassa \& Stappers 2004, 22-Grindlay et al.\ 2002, 
23- Becker et al.\ 2003, 24- D'Amico et al.\ 2002, 25- Nicastro et al.\ 2004, 26- Toscano et al.\ 1999b,
27- Stappers et al.\ 2003, 28- Toscano et al.\ 1999a.
 }
\tablenotetext{a}{For the globular cluster MSPs, the host cluster is given in parentheses.}
\tablenotetext{b}{Type of emission dominating in the X-ray range: thermal (th), magnetospheric (mag),
or emission from an unresolved shock (sh). The most uncertain cases are marked with `?'.}
\tablenotetext{c}{Orbital period for binary MSPs.}
\tablenotetext{d}{Transverse velocity.}
\tablenotetext{e}{Intrinsic spin-down power, corrected for the Shklovskii
effect and the effect of gravitational potential for globular cluster pulsars.}
\tablenotetext{f}{Distances estimated from the model for Galactic electron distribution by
Cordes \& Lazio (2002) unless indicated otherwise.}
\tablenotetext{g}{Photon index from fitting with one-component PL model (irrespective of fit quality),
characterizing the spectral hardness.}
\tablenotetext{h}{Best estimate for the unabsorbed X-ray luminosity in the 0.1--10 keV band.}
\tablenotetext{i}{$\eta = L_{\rm X}/\dot{E}$ is the total X-ray efficiency.}
\tablenotetext{j}{Distance measured from radio-timing parallax.}
\tablenotetext{k}{Since proper motion was not measured for this pulsar, we cannot correct $\dot{E}$ for the Shklovskii effect.}
\tablenotetext{l}{Double neutron star binary.}
\tablenotetext{m}{Distance estimated from observations of the white dwarf companion.}
\tablenotetext{n}{Triple system with a planet. The intrinsic $\dot{P}$ and $\dot{E}$ are constrained very poorly (see text).}
\tablenotetext{p}{Distance to the globular cluster.}
%\end{deluxetable}
\end{table}


\begin{thebibliography}

\bibitem[]{744}
Bailes, M., et al. 1994, 425, L41

\bibitem[]{747}
Bassa, C.\ G., \& Stappers, B.\ W. 2004, A\&A, 426, 1143

\bibitem[]{750}
Bassa, C., et al. 2004, ApJ, 609, 755

\bibitem[]{753}
Becker, W., \& Aschenbach, B. 2002, in Proc.\ 270th WE-Heraeus Seminar on
Neutron Stars, Pulsars, and Supernova Remnants, ed.\ W.\ Becker, H.\ Lesch,
\& J.\ Tr\"{u}mper (MPE Rep.\ 278; Garching: MPE), 64

\bibitem[]{758}
Becker, W., \& Pavlov, G.\ G. 2001, in The Century of Space Science,
ed.\ J.\ Bleeker, J.\ Geiss, \& M.\ Huber (Dordrecht: Kluwer), 721

\bibitem[]{762}
Becker, W., et al.\ 2003, ApJ, 594, 798

\bibitem[]{765}
Bogdanov, S., Grindlay, J.\ E., Heinke, C.\ O., Camilo, F., Freire, P.\ C.\ C., \&
Becker, W. 2006, ApJ, 646, 1104

\bibitem[]{769}
Bryden, G., Beichman, C.\ A., Rieke, G.\ H., Stansberry, J.\ A., Stapelfeldt, K.\ R.,
Trilling, D.\ E., Turner, N.\ J., \& Wolszczan, A. 2006, ApJ, 646, 1038

\bibitem[]{773}
Callanan, P.\ J., Garnavich, P.\ M., Coester, D. 1998, MNRAS, 298, 207

\bibitem[]{776}
Cash, W. 1979, ApJ, 228, 939 

\bibitem[]{794}
Chatterjee, S., Gaensler, B.\ M., Melatos, A., Brisken, W.\ F., \& Stappers, B.\ W. 2007, preprint (astro-ph/0703181)

\bibitem[]{779}
Cordes, J.\ M., \& Lazio, T.\ J. 2002, preprint (astro-ph/0207156)

\bibitem[]{800}
Cordes, J.\ M., \& Shannon, R.\ M. 2006, preprint (astro-ph/0605145)

\bibitem[]{782}
D'Amico, N., Possenti, A., Fici, L., Manchester, R.\ N., Lyne, A.\ G., Camilo, F.,
\& Sarkissian, J. 2002, ApJ, 570, L89

\bibitem[]{786}
Dickey, J.\ M., \& Lockman, F.\ J. 1990, ARA\&A, 28, 215

\bibitem[]{789}
Grindlay, J.\ E., Camilo, F., Heinke, C.\ O., Edmonds, P.\ D., Cohn, H., \& Lugger, P.
2001, ApJ, 581, 470

\bibitem[]{793}
Hotan, A.\ W., Bailes, M., \& Ord, S.\ M. 2006, MNRAS, 369, 1502

\bibitem[]{796}
Kargaltsev, O., Pavlov, G.\ G., \& Garmire, G.\ P. 2006a, ApJ, 636, 406

\bibitem[]{799}
Kargaltsev, O., Pavlov, G.\ G., \& Garmire, G.\ P. 2006b, ApJ, 646, 1139

\bibitem[]{802}
Konacki, M., \& Wolszczan, A. 2003, ApJ, 591, L147

\bibitem[]{805}
Kramer, M., et al. 2006, Science, 314, 97

\bibitem[]{808}
Lange, Ch., Camilo, F., Wex, N., Kramer, M., Backer, D.\ C., Lyne, A.\ G.,
\& Doroshenko, O. 2001, MNRAS, 326, 274

\bibitem[]{812}
Lommen, A.\ N., Kipphorn, R.\ A., Nice, D.\ J., Splaver, E.\ M., Stairs, I.\ H., \&
Backer, D.\ C. 2006, ApJ, 642, 1012

\bibitem[]{837}
Lorimer, D.\ R. 2005, Living Rev.\ Relativity, 8, 7 (cited on 2007 March 20)

\bibitem[]{816}
Lyne, A.\ G., et al. 2004, Science, 303, 1153

\bibitem[]{819}
Miller, M.\ C., \& Hamilton, D.\ P. 2001, ApJ, 550, 863

\bibitem[]{822}
Navarro, J., de Bruyn, A.\ G., Frail, D.\ A., Kulkarni, S.\ R., \& Lyne, A.\ G. 1995,
ApJ, 455, L55

\bibitem[]{826}
Nicastro, L., Cusumano, G., L\"{o}hmer, O., Kramer, M., Kuiper, L.,
Hermsen, W., Mineo, T., \& Becker, W. 2004, A\&A, 413, 1065

\bibitem[]{830}
Nice, D.\ J., Splaver, E.\ M., Stairs, I.\ H., L\"{o}hmer, O., Jessner, A., Cramer, M.,
Cordes, J.\ M. 2005, ApJ, 634, 1242

\bibitem[]{}
Nousek, J.\ A., \& Shue, D.\ R. 1989, ApJ, 342, 1207

\bibitem[]{834}
Phinney, E.\ S., \& Hansen, B.\ M.\ S. 1993, in 
ASP Conf.\ Ser.\ 36, Planets around Pulsars, ed.\ J.\ A.\ Phillips, 
S.\ E.\ Thorsett, \& S.\ R.\ Kulkarni (San Francisco: ASP), 371

\bibitem[]{839}
Podsiadlowski, P. 1993, in ASP Conf.\ Ser.\ 36, Planets around Pulsars, 
ed.\ J.\ A.\ Phillips, S.\ E.\ Thorsett, \& S.\ R.\ Kulkarni 
(San Francisco: ASP), 149

\bibitem[]{844}
Shklovskii, I.\ S. 1970, Sov.\ Astron., 13, 562

\bibitem[]{847}
Sigurdsson, S., \& Thorsett, S.\ E. 2005, in Binary Radio Pulsars,
ASP Conf.\ Ser., v.328, ed.\ F.\ A.\ Rasio \& I.\ H.\ Stairs (San Francisco: ASP), p.213

\bibitem[]{851}
Stairs, I.\ H., Thorsett, S.\ E., Taylor, J.\ H., \& Wolszczan, A. 2002, ApJ, 581, 501

\bibitem[]{854}
Stappers, B.\ W., Gaensler, B.\ M., Kaspi, V.\ M., van der Klis, M., \& Lewin, W.\ H.\ G. 2003,
Science, 299, 1372

\bibitem[]{858}
Taylor, J., \& Cordes, J.  1993, ApJ, 411, 674

\bibitem[]{861}
Thorsett, S.\ E., Arzoumanian, Z., Camilo, F., \& Lyne, A.\ G. 1999, ApJ, 523, 763

\bibitem[]{864}
Toscano, M., Britton, M.\ C., Manchester, R.\ N., Bailes, M., Sandhu, J.\ S.,
Kulkarni, S.\ R., \& Anderson, S.\ B. 1999a, ApJ, 523, L171

\bibitem[]{868}
Toscano, M., Sandhu, J.\ S., Bailes, M., Manchester, R.\ N., Britton, M.\ C.,
Kulkarni, S.\ R., Anderson, S.\ B., \& Stappers, B.\ W. 1999b, MNRAS, 307, 925

\bibitem[]{872}
Webb, N.\ A., Olive, J.-F., \& Barret, D. 2004a, A\&A, 417, 181

\bibitem[]{875}
Webb, N.\ A., Olive, J.-F., Barret, D., Kramer, M., Cognard, I., L\"{o}hmer, O.
2004b, A\&A, 419, 269

\bibitem[]{879}
Wilms, J., Allen, A., \& McCray, R. 2000, ApJ, 542, 914

\bibitem[]{882}
Wolszczan, A. 1990, IAU Circ.\ 5073

\bibitem[]{885}
Wolszczan, A., \& Frail, D.\ A. 1992, Nature, 355, 145

%\bibitem[]{888}
%Wolszczan, A., \& Konacki, M. 2005, in Planet Formation and Detection,
%Aspen 2005 Winter Conference on Astrophysics,
%http://www.astro.northwestern.edu/AspenW05/program.html 

\bibitem[]{893}
Wolszczan, A., et al. 2000, ApJ, 528, 907

\bibitem[]{920}
Xilouris, K.\ M., et al. 1998, ApJ, 501, 286

\bibitem[]{896}
Zavlin, V.\ E. 2006a, ApJ, 2006, ApJ, 638, 951

\bibitem[]{899}
Zavlin, V.\ E. 2006b, Astrophys.\ Space Sci., in press
(astro-ph/0608210)

\bibitem[]{903}
Zavlin, V.\ E., \& Pavlov, G.\ G. 2004, ApJ, ApJ, 616, 452

\bibitem[]{906}
Zavlin, V.\ E., Shibanov, Y.\ A., \& Pavlov, G.\ G. 1995, Astron.\ Lett.,
21, 149

\bibitem[]{910}
Zavlin, V.\ E., Pavlov, G.\ G., \& Shibanov, Y.\ A. 1996, A\&A, 315, 141

\bibitem[]{913}
Zavlin, V.\ E., Pavlov, G.\ G., Sanwal, D., Manchester, R.\ N., Tr\"umper, J.,
Halpern, J.\ P., \& Becker, W. 2002, ApJ, 569, 894

\bibitem[]{917}
Zhang, B., Sanwal, D., \& Pavlov, G.\ G., 2005, ApJ, 624, L109

\end{thebibliography}
\end{document}